\documentstyle[epsfig]{aipproc}

\begin{document}
\title{Observing Air Showers \\ from Cosmic Superluminal Particles}

\author{Luis Gonzalez-Mestres$^{*,\dagger}$}
\address{$^*$Laboratoire
de Physique Corpusculaire, Coll\`ege de France, 75231 Paris 
Cedex 05, France\\
$^{\dagger}$L.A.P.P., CNRS-IN2P3,
B.P. 110, 74941 Annecy-le-Vieux Cedex, France}

\maketitle

\begin{abstract}
The Poincar\'e relativity principle 
has been tested at low energy with great accuracy, but
its extrapolation to very high-energy phenomena is much less well established.
Lorentz symmetry can be broken at Planck scale due to 
the renormalization of gravity or to some deeper 
structure of matter:
we expect such a breaking to be a very high
energy and very short distance phenomenon.
If textbook special relativity is only an approximate property of the
equations describing a sector of matter above some critical distance scale,
an absolute local frame (the "vacuum rest frame", VRF) can possibly be found and
superluminal sectors of matter may exist related to new degrees of freedom
not yet discovered experimentally. The new superluminal particles
({\bf "superbradyons"}, i.e. bradyons with superluminal critical speed) would
have positive mass and energy, and behave kinematically
like "ordinary" particles (those with critical speed in vacuum equal to
$c$ , the
speed of light) apart from the difference in critical speed
(we expect $c_i~\gg ~c$ , where $c_i$ is the critical
speed of a superluminal sector). They may be the ultimate building blocks
of matter. At speed $v~ >~ c$ ,
they are expected to release "Cherenkov" radiation ("ordinary" particles) in
vacuum. Superluminal particles could provide most of the cosmic
(dark) matter and produce very high-energy cosmic rays.
We discuss: a) the possible
relevance of superluminal matter to the composition, sources and spectra of
high-energy cosmic rays; b) signatures and experiments allowing to possibly
explore such effects. Very large volume and unprecedented background
rejection ability are crucial requirements for any
detector devoted to the search for cosmic superbradyons. Future
cosmic-ray experiments
using air-shower detectors (especially from space) 
naturally fulfil both requirements.
\end{abstract}

{\it ~~~"The impossibility to disclose experimentally the absolute motion
of the earth seems to be a general law of Nature"
\vskip 1mm
H. Poincar\'e
\vskip 4mm
"The interpretation of geometry advocated here cannot be directly applied
to submolecular spaces... it might turn out that such an extrapolation is
just as incorrect as an extension of the concept of temperature to particles
of a solid of molecular dimensions"
\vskip 1mm
A. Einstein}

\section*{Relativity, matter and critical speeds}
 
If Lorentz symmetry is viewed as a dynamical property of the motion
equations, no reference to absolute
properties of space and time is required
\cite{Gon95a}. In a two-dimensional
galilean space-time,
the equation:
\equation
\alpha ~\partial ^2\phi /\partial t^2~-~\partial ^2\phi /\partial x^2 = F(\phi )
\endequation

\noindent
with $\alpha$ = $1/c_o^2$ and $c_o$ = critical
speed, remains unchanged under "Lorentz" transformations leaving
invariant the squared
interval:
\equation
ds^2 = dx^2 - c_o^2 dt^2
\endequation

\noindent
so that matter made with solutions of equation (1)
would feel a relativistic space-time even if the real space-time is actually
galilean and if an absolute rest frame exists in the
underlying dynamics beyond the wave equation.
A well-known example is provided by the solitons of the sine-Gordon equation,
obtained taking in (1):
\equation
F(\phi )~~ = ~~-~(\omega /c_o)^2~sin~\phi
\endequation

\noindent
where $\omega $ is a characteristic frequency of the dynamical system.
A two-dimensional universe made of sine-Gordon solitons plunged
in a galilean world would behave like a two-dimensional minkowskian
world with the laws of special relativity.
Information on any absolute rest frame would be lost by the solitons, as
if the Poincar\'e relativity principle (see \cite{Po95} to \cite{Po05} for the
genesis and evolution of this deep concept) 
were indeed a law of Nature, even if
actually the
basic equation derives from a galilean world with an
absolute rest frame
(a system built "on top of a table", with $c_o ~\ll ~c$). 
The actual structure of space and time
can only be found by going beyond the wave equation
to deeper levels of resolution, similar to the way
high-energy accelerator experiments explore the inner structure of
"elementary" particles (but cosmic rays have the highest attainable 
energies). 

At this stage, two crucial questions arise:
a) is $c$ (the speed of light) the only critical speed in vacuum, are
there particles with a critical speed different from that of light?
\cite{Gon95a,Gon95b}; 
b) can the ultimate building blocks of matter be superluminal?
\cite{Gon97a,Gon97b}.
These questions make sense, as: a) in a
perfectly transparent crystal it is possible to identify
at least two critical speeds, those of light and
sound, and light can interact with phonons; b) 
the potential approach to lattice dynamics in solid-sate physics
is precisely the form of electromagnetism
in the limit $c_s~c^{-1}~\rightarrow ~0$ , where $c_s$ is the speed of sound. 
Superluminal
sectors of matter can be consistently
generated \cite{Gon95a,Gon96a}, 
with the conservative choice of leaving the Planck constant
unchanged, replacing in the Klein-Gordon equation the
speed of light by a new critical speed $c_i$ $\gg $ $c$
(the subscript $i$ stands for the $i$-th superluminal sector). All
standard kinematical concepts and
formulas \cite{Sch61} remain correct, leading to particles with
positive mass and energy which are not tachyons.
We shall call them {\bf superbradyons} as, according to standard
vocabulary \cite{Rec78}, they are bradyons with superluminal critical
speed in vacuum. The energy $E$ and momentum $p$
of a superluminal particle of mass $m$ and critical speed $c_i$
will be given by the generalized relativistic equations:
\begin{eqnarray}
p~=~m~v~(1~-~v^2c_i^{-2})^{-1/2}~ \\
~E~=~m~c_i^2~(1~-~v^2c_i^{-2})^{-1/2} \\
E_{rest}~~=~~m~c_i^2~~~~~~~~~~~~~~~
\end{eqnarray}

\noindent
where $v$ is the speed and $E_{rest}$ the rest energy.
Energy and momentum conservation will in principle not be
spoiled by the existence of several critical speeds in vacuum:
conservation laws will as usual hold for phenomena leaving the vacuum
unchanged. Each superluminal sector will have its own Lorentz invariance
with $c_i$ defining the metric, and is expected to generate a sectorial
"gravity".
Interactions between two different
sectors will break both Lorentz invariances. Lorentz
invariance
for all sectors simultaneously will at best be explicit
(i.e. exhibiting the diagonal sectorial Lorentz metric) in a single
inertial frame ({\bf the vacuum rest frame}, VRF, i.e. the "absolute" rest
frame). In our approach, the Michelson-Morley result is not incompatible
with the existence of some "ether" as
suggested by recent results in particle physics: if
the vacuum is a material medium where fields
and order parameters can condense, it may well
have a local rest frame whose identification would be prevented by the
sectorial Lorentz symmetries in the low-momentum limit (where different 
sectors do not mix and the sectorial Lorentz symmetries become exact laws,
so that each sector feels a "Poincar\'e relativity principle"). 

If superluminal particles couple
weakly to ordinary matter, their effect on the ordinary sector will occur at
very high energy and short distance \cite{Gon97c}, far from
the domain of successful
conventional tests of Lorentz invariance
\cite{Lam,Hills}.
In particular, superbradyons naturally escape the
constraints on the critical speed derived in some specific models
\cite{Col,Glas} based on the $TH\epsilon \mu $ approach \cite{Will} , as their
mixing with the ordinary sector is expected to be strongly energy-dependent
\cite{Gon97b,Gon97d}. 
High-energy experiments can therefore
open new windows in this field.
Finding some track of a superluminal sector (e.g. through
violations of Lorentz invariance in the ordinary sector
or by direct detection of a superluminal particle) may
be the only way to experimentally discover the VRF.
Superluminal particles lead to consistent cosmological models
\cite{Gon95b,Gon97a,Gon96a}, where they may well provide
most of the cosmic (dark) matter \cite{Gon96b}. Although recent criticism
to this suggestion has been emitted in a specific model
on the grounds of gravitation theory \cite{Konst},
the framework used is
crucially different
from the multi-graviton approach suggested in our papers where we propose
(e.g. \cite{Gon95a,Gon96a}) that each dynamical (ordinary or superluminal)
sector generates its own gravitation associated to the sectorial Lorentz 
symmetry and couplings between different "gravitons" are expected to be weak.
Superbradyons can be the ultimate building blocks from
which superstrings would be made and a "pre-Big Bang" 
cosmology would emerge. Nonlocality at Planck scale
would then be an approximation to this dynamics
in the limit $c~c_i^{-1}~\rightarrow ~0$ , where superluminal signals undergo
apparent
"instantaneous" 
propagation similar to electromagnetic interactions described by
a potential model of lattice dynamics
in solid state physics.

\section*{Implications for high-energy cosmic rays}

Accelerator experiments at future machines (LHC, VLHC...) can be a way
to search for superluminal particles
\cite{Gon97c,Gon97d}. However, this approach is limited by
the attainable energies, luminosities, signatures and low-background levels. 
Although the investigation at accelerators provides unique chances and
must be carried on, it will only cover a small domain of the allowed parameters
for superluminal sectors of matter. Cosmic-ray experiments 
are not limited in energy and naturally provide very low background levels:
they therefore allow for a more general and, on dynamical grounds, 
better adapted exploration. It must also be realized that, if the Poincar\'e
relativity principle is violated, a $1~TeV$ particle cannot be turned into 
a $10^{20}~eV$ particle of the same kind by a Lorentz transformation, and
collider events cannot be made equivalent to cosmic-ray events.
 
The highest observed cosmic-ray energies (up to 3.10$^{20}~eV$) 
are closer to Planck scale ($\approx ~10^{28}~eV$) than to
electroweak 
scale ($\approx ~10^{11}~eV$): therefore, if Lorentz symmetry is violated,
the study of the highest-energy cosmic rays
provides a unique microscope directly focused on Planck scale
\cite{Gon97a,Gon97b,Gon97e}. The search for very rare events due to
superluminal particles in AUGER, AMANDA, OWL, AIRWATCH FROM SPACE...
can be a crucial ingredient of this unprecedented
investigation \cite{Gon97c,Gon97d,Gon97f,Gon97g}. In what follows we assume 
that the earth is not moving at relativistic speed with respect to the local
vacuum rest frame.

\subsection*{Superluminal kinematics}
The kinematical properties and Lorentz transformations
of high-energy superluminal particles have been
discussed elsewhere \cite{Gon97c}.
If an absolute rest frame exists, Lorentz contraction is a real physical
phenomenon and is governed by the factor
$\gamma _i^{-1}~=~(1~-~v^2c_i^{-2})^{1/2}$
for the $i$-th superluminal sector,
so that there is no Lorentz singularity when a superluminal particle crosses
the speed value $v~=~c$ in a frame measured by ordinary
matter. Similarly, if superbradyons have any coupling to the
electromagnetic field (adding in the standard way the electromagnetic
four-potential to the superluminal four-momentum to build the covariant
derivative in the VRF),
we expect the magnetic force to be proportional to
$v~c_i^{-1}$ instead of $v~c^{-1}$~.
Contrary to tachyons, superbradyons
can emit "Cherenkov" radiation (i.e. particles with lower
critical speed) in vacuum.
If $c_i~\gg ~10^3~c$ , and if the VRF is close to that defined
requiring isotropy of cosmic microwave background radiation,
high-energy superluminal particles will be seen on earth
as traveling mainly at speed $v~\approx ~10^3~c$ , as can be seen from
the following analysis.
Since we expect to measure the energy
of superluminal particles through interactions with detectors made of
"ordinary" particles,
we can define, in the rest frame of an "ordinary" particle moving at speed
${\vec {\mathbf V}}$ with respect to the VRF, the energy and
momentum of a superluminal particle to be the Lorentz-tranformed
of its VRF energy and
momentum taking $c$ as the critical speed parameter for the Lorentz
transformation. Then, the mass of the superluminal particle will depend
on the inertial frame. The energy $E'_i$ and momentum 
${\vec {\mathbf p}}{\mathbf '_i}$
of 
the superluminal particle $i$ (belonging to the $i$-th superluminal
sector and with energy $E_i$ and momentum ${\vec{\mathbf p}}{\mathbf _i}$
in the VRF) in the new rest
frame, as measured by ordinary matter from energy and momentum
conservation (e.g. in decays of superluminal particles into ordinary
ones), will be:
\begin{eqnarray}
E'_i~~=~~(E_i~-~{\vec {\mathbf V}}.{\vec {\mathbf p}}{\mathbf _i})~
(1~-~V^2c^{-2})^{-1/2} ~~~~~~~~~~~\\
{\vec {\mathbf p}}{\mathbf '_i}~~=~~
{\vec {\mathbf p}}{\mathbf '_{i,L}}~+~
{\vec {\mathbf p}}{\mathbf '_{i,\perp }}~~~~~~~~~~~~~~~~~~~~~~~~~~~~~~~~~~~~ \\
~~~{\vec {\mathbf p}}{\mathbf '_{i,L}}~~=~~
({\vec {\mathbf p}}{\mathbf _{i,L}}~
-~E_i~c^{-2}~{\vec {\mathbf V}})~(1~-~V^2c^{-2})^{-1/2}~ \\
{\vec {\mathbf p}}{\mathbf '_{i,\perp }}~~=~~
{\vec {\mathbf p}}{\mathbf _{i,\perp }}~~~~~~~~
~~~~~~~~~~~~~~~~~~~~~~~~~~~~~~~~~~~~
\end{eqnarray}
where  ${\vec {\mathbf p}}{\mathbf _{i,L}}
~=~V^{-2}~
({\vec {\mathbf V}}.{\vec {\mathbf p}}{\mathbf _i})~{\vec {\mathbf V}}$~ ,~
${\vec {\mathbf p}}{\mathbf _{i,\perp }}~~=~~
{\vec{\mathbf p}}{\mathbf _i}~-~{\vec {\mathbf p}}{\mathbf _{i,L}}$
and similarly for the longitudinal and transverse components
of ${\vec {\mathbf p}}{\mathbf '_i}$ .
We are thus led to consider the effective squared mass:
\equation
M_{i,c}^2~~=~~c^{-4}~(E_i^2~-~c^2p_i^2)~~=~~
m_i^2c^{-4}c_i^4~+~c^{-2}(c^{-2}c_i^2~-~1)~p_i^2
\endequation
which depends on the VRF momentum of the particle.
$m_i$ is the invariant mass of particle $i$ , as seen by matter from the
$i$-th
superluminal sector (i.e. with critical speed in vacuum = $c_i$).
While "ordinary" transformation laws of energy and momentum are not singular,
even for a superluminal particle, the situation is different for
the transformation of a superluminal
speed, as will be seen below. Furthermore,
if the superluminal particle has velocity ${\vec {\mathbf v}}{\mathbf _i}~=~
{\vec {\mathbf V}}$ in the VRF, so that it is at rest in the new
inertial frame, we would naively expect a vanishing momentum,
${\vec {\mathbf p}}{\mathbf '_i}~=~ {\mathbf 0}$ . Instead, we get:
\equation
{\vec {\mathbf p}}{\mathbf '_i}~~=
~~-~{\vec{\mathbf p}}{\mathbf _i}~(c^{-2}c_i^2~-~1)~(1~-~V^2c^{-2})^{-1/2}
\endequation
and $p'_i~\gg ~p_i$ , although $p'_ic~\ll ~E'_i$ if $V~\ll ~c$ .
This reflects the non-covariant character of the 4-momentum of particle $i$
under "ordinary" Lorentz transformations. Thus, even if the directional
effect is small
in realistic situations (f.i. on earth),
the decay of a superluminal particle at rest into
ordinary particles will not lead to an exactly vanishing total momentum if the
inertial frame is different from the VRF.

In the rest frame of an "ordinary" particle moving with speed
${\vec {\mathbf V}}$
with respect to the VRF, we can estimate the speed
${\vec {\mathbf v}}{\mathbf '_i}$ 
of the previous particle $i$ writing:
\equation
{\vec {\mathbf v}}{\mathbf _i}~~=~~{\vec {\mathbf v}}
{\mathbf _{i,L}}~+~{\vec {\mathbf v}}{\mathbf _{i,\perp }}
\endequation
where ${\vec {\mathbf v}}
{\mathbf _{i,L}}
= V^{-2}
({\vec {\mathbf V}}.{\vec {\mathbf v}}{\mathbf _i})~{\vec {\mathbf V}}$~ ,
and similarly for the longitudinal and transverse components
of ${\vec {\mathbf v}}{\mathbf '_i}$ . Then, the transformation law is:
\begin{eqnarray}
{\vec {\mathbf v}}
{\mathbf '_{i,L}}~~=~~({\vec {\mathbf v}}{\mathbf _{i,L}}~
-~{\vec {\mathbf V}})~
(1~-~{\vec {\mathbf v}}{\mathbf _i}.{\vec {\mathbf V}}~ c^{-2})^{-1}~~~~~~~~~~~~ \\
~~~~~~~~~{\vec {\mathbf v}}{\mathbf '_{i,\perp }}
~~=~~{\vec {\mathbf v}}{\mathbf _{i,\perp }}
~(1~-~V^2c^{-2})^{1/2}~(1~-~
{\vec {\mathbf v}}{\mathbf _i}. {\vec {\mathbf V}}~c^{-2})^{-1}~~~
\end{eqnarray}
leading to singularities at ${\vec {\mathbf v}}{\mathbf _i}~~=~~c^2$
which correspond to a change in the arrow of time (due to the distorsion 
generated by the Lorentz
transformation of space-time)
as seen by ordinary matter
traveling at speed ${\vec {\mathbf V}}$ with respect to the VRF.

\subsection*{Experimental implications}
 
At $v_{i,L}~>~c^2V^{-1}$ , a superluminal particle moving forward in time in
the VRF will appear as moving backward in time to
an observer made of ordinary matter and moving at speed ${\vec {\mathbf V}}$
in the same frame. On earth, taking $V~\approx ~10^{-3}~c$
(if the VRF is close to that suggested by cosmic background
radiation, e.g. \cite{Pee}), the apparent
reversal of the time arrow will occur mainly at
$v_i~\approx ~ 10^{3}~c~.$ If $c_i~\gg ~10^3~c$~,
phenomena related to propagation backward in time of produced superluminal
particles may be observable in future
accelerator experiments slightly above the
production threshold. In a typical event where a pair of superluminal particles
would be produced, we expect in most cases that one of the superluminal
particles propagates forward in time and the other one propagates backward.
As previously stressed, the infinite velocity (value
of $v'_i$)
associated to the point of time reversal does not, according to (7) and (9), 
correspond to infinite
values of energy and momentum.
The backward propagation in time, as observed by devices which are not at
rest in the VRF, is not really physical
(the arrow of time is well defined in the VRF for all
physical processes) and does not
correspond to any real violation of causality.
The apparent reversal of the time arrow for superluminal particles at
${\vec {\mathbf v}}{\mathbf _i}.{\vec {\mathbf V}}~>~c^2$ would
be a consequence of
the bias of the laboratory time measurement due to our motion
with respect to the absolute rest frame.
The distribution and properties
of such superluminal events, in an accelerator experiment 
or in a large-volume cosmic-ray
detector,
would obviously be in correlation with the direction and
speed of the laboratory's motion with respect to the VRF. 
It would provide fundamental cosmological information, 
complementary to informations
on "ordinary" matter provided by measurements of the cosmic
microwave background.

From (14) and (15), we also notice that, for $V~\ll ~c$ and
${\vec {\mathbf v}}{\mathbf _i}. {\vec {\mathbf V}}~\gg ~c^2$ , the speed
${\vec {\mathbf v}}{\mathbf {'_i}}$ tends to the limit
${\vec {\mathbf v}}{\mathbf _i^{\infty }}$ , where:
\equation
{\vec {\mathbf v}}{\mathbf _i^{\infty }}~
({\vec {\mathbf v}}{\mathbf _i})~~=~~-~{\vec {\mathbf v}}{\mathbf _i}~c^2~
({\vec {\mathbf v}}{\mathbf _i}.{\vec {\mathbf V}})^{-1}
\endequation
which sets a universal high-energy limit, independent of $c_i$ ,
to the speed of superluminal particles as
measured by ordinary matter in an inertial rest frame other than the 
VRF. This limit is not isotropic, and depends on the angle between
the speeds ${\vec {\mathbf v}}{\mathbf _i}$ and ${\vec {\mathbf V}}$ .
A typical order of magnitude for
${\vec {\mathbf v}}{\mathbf _i^{\infty }}$ on earth is
${\vec {\mathbf v}}{\mathbf _i^{\infty }}~\approx 10^3~c$ if
the VRF is close to
that suggested by cosmic background radiation.
If $C$ is the
highest critical speed in vacuum, infinite speed and reversal of the arrow
of time occur only in frames moving with respect to the VRF 
at speed $V~\geq ~c^2C^{-1}$ . Finite critical speeds
of superluminal sectors, as measured by ordinary matter in frames
moving at $V~\neq ~0$ , are anisotropic. Therefore, directional detection
of superluminal particles would allow to directly identify the VRF 
and even to check whether it can be defined 
consistently, simultaneously for all dynamical sectors. If a universal, local
VRF cannot be defined, translational and rotational modes may appear between
different kinds of matter generating significant cosmological effects (e.g.
a cosmic rotation axis for "ordinary" matter).
 
A superbradyon moving with velocity ${\vec {\mathbf v}}{\mathbf _i}$
with
respect to the VRF,
and emitted by an astrophysical object, can reach an observer
moving with laboratory
speed  ${\vec {\mathbf V}}$ in the VRF at a time, as
measured by the observer, previous to the emission time. 
This remarkable astronomical phenomenon will
happen if ${\vec {\mathbf v}}{\mathbf _i}.{\vec {\mathbf V}}~>~c^2$ , and the 
emitted particle will be seen to evolve backward in time (but it evolves
forward in time in the VRF, so that again the reversal of the time
arrow is not really a physical phenomenon).
If they interact several times with
the detector,
superbradyons can
be a directional probe preceding the detailed observation of
astrophysical phenomena, such as explosions releasing
simultaneously neutrinos, photons and superluminal particles
(although causality is preserved in the VRF).
For a high-speed
superluminal cosmic ray with critical speed $c_i~\gg ~c$ ,
the momentum, as measured in the laboratory,
does not provide directional
information on the source, but on the VRF.
Velocity provides directional information on the source,
but can be measured only if the
particle interacts several times with the detector, which is far from
guaranteed, or if the superluminal particle is associated to a collective
phenomenon involving several sectors of matter and 
emitting also photons or neutrinos simultaneously.
In the most favourable case,
directional detection of high-speed
superluminal particles in a very large 
detector would allow
to trigger a dedicated astrophysical observation in the direction of the sky
determined by the velocity of the superluminal particle(s).
If $d$ is the distance between the observer and the astrophysical object,
and $\Delta t$ the time delay between the detection of the superluminal
particle(s) and that of photons and neutrinos,
we have: $d~\simeq ~c~\Delta t$~.

Annihilation of pairs of superluminal particles into ordinary ones can
release very large kinetic energies and provide a new
source of high-energy cosmic rays.
Decays of superluminal particles may play a similar role.
Collisions (especially, inelastic with very large energy
transfer) of high-energy superluminal particles with
extra-terrestrial ordinary matter may also yield high-energy
ordinary cosmic rays. Pairs of slow superluminal particles can also
annihilate into particles of another superluminal sector
with lower $c_i$ , converting most of the rest energies into a
large amount of kinetic energy.
Superluminal particles moving at $v_i~>~c$ can release anywhere "Cherenkov"
radiation in vacuum, i.e.
spontaneous emission of particles of a
lower critical speed $c_j$ (for $v_i~>~c_j$) including ordinary ones,
providing a new source of (superluminal or ordinary)
high-energy cosmic rays.
High-energy superluminal particles
can directly reach the earth and undergo
collisions inside the atmosphere, producing many secondaries like
ordinary cosmic rays. They can also interact with the rock or
with water near some underground or underwater detector,
coming from the atmosphere or after having crossed the earth,
and producing clear signatures.
Contrary to neutrinos, whose flux is strongly attenuated by the
earth at
energies above $10^6$ $GeV$ , superluminal particles will in
principle not be stopped by earth at these energies.
In inelastic collisions, high-energy superluminal primaries can
transfer most of their energy to ordinary particles.
Even with a very weak interaction probability,
and assuming that the superluminal primary does not produce ionization,
the rate for superluminal cosmic ray events can be observable
if we are surrounded by important concentrations of superluminal
matter, which is possible in suitable cosmologies \cite{Gon97a}. 
Atypical ionization properties would further 
enhance background rejection, but ionization can be in contradiction
with the requirement  
of very weak coupling to ordinary matter unless the coupling is 
energy-dependent. 

The possibility that superluminal matter exists, and that it plays
nowadays an important role in our Universe,
should be kept in mind when addressing the two basic questions
raised by the analysis of any cosmic-ray event:
a) the nature and properties of the cosmic-ray primary; b)
the identification (nature and position) of the source of the cosmic ray.
If the primary
is a superluminal particle, it will escape conventional criteria
for particle identification
and most likely produce a specific signature
(e.g. in inelastic collisions) different from those of ordinary
primaries.
Like neutrino events, in the absence of
ionization we
may expect the event to start anywhere inside the detector.
Unlike very high-energy neutrino events,
events created by superluminal primaries can
originate from a particle having crossed the earth.
An incoming, relativistic superluminal particle with momentum $p$ and
energy $E_{in} \simeq p_i~ c_i $ in the VRF, 
hitting an ordinary particle at rest,
can, for instance, release most of its energy into
two ordinary particles with
momenta (in the VRF)
close to $p_{max}~=~1/2~p_i~c_i~c^{-1}$ and oriented back
to back in such a way
that the two momenta almost cancel.
Then, an energy $E_R \simeq E_{in} $
would be transferred to ordinary secondaries. More generally,
we can expect several jets in a configuration with 
very small total momentum as compared to
$c^{-1}$ times the total energy, or a basically isotropic event.
Corrections due to the earth motion
must be applied (see previous Section) before defining the expected
event configuration in laboratory or air-shower experiments, but the basic 
trends just described remain.
At very high energy, such events
would be easy to identify in large volume detectors, even at very small rate.
If the
source is superluminal, it can be located anywhere
(and even be a free particle in the case of "Cherenkov"
emission) and will not necessarily be at the same place as
conventional sources
of ordinary cosmic rays. High-energy cosmic-ray events
originating form superluminal sources
will provide hints on the location of such
sources and be possibly the only way to observe them.
The energy dependence of the events should be taken into account.

At very high energies,
the Greisen-Zatsepin-Kuzmin (GZK)
cutoff \cite{Grei,Zats}
does not in principle hold for
cosmic-ray events originating from superluminal matter:
this is obvious if the primaries are superluminal particles
that we expect to interact very weakly with the cosmic microwave
background,
but applies also in practice to ordinary primaries as we do not expect them to
be produced at the locations of ordinary sources and there is no upper
bound to their energy around $100~EeV$.
Besides "Cherenkov" deceleration, a superluminal cosmic
background radiation may exist
and generate its own GZK
cutoffs for the superluminal sectors. However, if there are large amounts
of superluminal matter around us, they can be the main superluminal source
of cosmic rays reaching the earth.
To date, there is
no well-established interpretation of the
highest-energy cosmic-ray events.
Primaries (ordinary or superluminal)
originating from superluminal particles are acceptable candidates
and can possibly escape several problems
(event configuration, source location, energy dependence...)
faced by cosmic rays produced at ordinary sources.

\section*{Potentialities of Air-Shower detectors}

Since the discovery of superluminal matter would be an
unprecedented event in the history of Physics, and we do not know at what
energy scale it would manifest itself,
direct detection of cosmic superluminal particles (CSL) deserves special
consideration having in mind the exceptional potentialities
of future cosmic-ray detectors.
As we expect a very weak coupling between superluminal and
"ordinary" matter, except possibly at Planck scale, it is crucial to be able
to cover an unusually large target volume. 
If the coupling increases with energy, it can compensate the possible fall with energy of the CSL flux and make the highest-energy experiments especially
adapted to the search for CSL.
Future air-shower
detectors devoted to the highest-energy cosmic rays
will observe the largest target volumes ever reached
in a particle physics experiment (especially in the case of
satellite-based programs such as OWL or AIRWATCH FROM SPACE). 
Due to the energies they are
able to cover, and considering the possibility that 
Lorentz symmetry be violated at Planck scale,
such experiments are as sensitive to
phenomena generated by Planck-scale physics as any
possible particle physics experiment can be.

To possibly observe CSL, 
background rejection must be unprecedentedly powerful. This would be the
case for ultra-high energy events generated by CSL.
As previously stressed, the ratio $E_{in} \simeq p_i~ c_i $ (in the VRF) 
provides a unique event profile: since the total momentum of the
produced ordinary particles is very small as compared to the total available
energy (using $c$ as the conversion factor), the event cannot have the usual,
sharply forward-peaked shape of showers produced by "ordinary" cosmic rays.
Instead, it can be made of two or more (broad) jets, or be basically isotropic.
No "ordinary" ultra-high energy particle can produce such an event shape. The
discussion remains valid in any reference frame moving at low speed with respect
to the VRF, with the corrections discussed previously.
Therefore, air-shower detectors should basically
look for events originating at any
depth in the atmosphere (like neutrino-induced events)
but which, unlike neutrino
events where a single elementary
particle gets part of the incoming neutrino momentum
and subsequently produces a conventional shower profile, do not present a
single privileged direction for the produced particles
and have instead a tendency to be isotropic. Furthermore, if the
earth moves at a speed $\approx 10^{-3}~c$ with respect to the VRF 
and $c~c_i^{-1}~\ll~10^{-3}$,
the total momentum of the produced particles must in most events cancel with
$\approx 10^{-3}$ precision as compared to the total energy, up to fluctuations
due to unobserved neutrals and to measurement uncertainties.

\end{document}